# Powering Quantum Computation with Quantum Batteries

Yaniv Kurman,[1] Kieran Hymas,[1] Arkady Fedorov,[2] William J. Munro,[3] and James Quach[1,*]

[1]*Commonwealth Scientific and Industrial Research Organisation (CSIRO), Clayton, Victoria 3168, Australia*

[2]*School of Mathematics and Physics, The University of Queensland, Brisbane, QLD 4072, Australia*

[3]*Okinawa Institute of Science and Technology Graduate University, Onna-son, Kunigami-gun, Okinawa, 904-0495, Japan*

Executing quantum logic in cryogenic quantum computers requires a continuous energy supply from room-temperature control electronics. This dependence on external energy sources creates scalability limitations due to control channel density and heat dissipation. Here, we propose quantum batteries (QBs) as *intrinsic* quantum energy sources for quantum computation, enabling the thermodynamic limit of zero dissipation for unitary gates. Unlike classical power sources, QBs maintain quantum coherence with their load—a property that, while theoretically studied, remains unexploited in practical quantum technologies. We demonstrate that initializing a bosonic QB in a Fock state can supply the energy required for arbitrary unitary gates regardless of the circuit's depth, via the recycling of precharged energy. Crucially, allowing QB-qubit entanglement during computation lowers the QB's initial energy requirements below established energy-fidelity bounds. This scheme facilitates a universal gate set controlled by a single parameter per qubit: its resonant frequency. The relative detuning of each qubit from the QB's resonant frequency qualitatively gives rise to two gate types: off resonance and around resonance. The former facilitates dispersive gates that allow multiqubit parity probing while the latter enables energy exchange between the QB and the qubits, driving both population transfer and entanglement generation. This mechanism utilizes the all-to-all connectivity of the shared-resonator architecture to go beyond the standard single- and two-qubit native gates of current platforms with multiqubit gate timescales of few $\pi/g$, where $g$ is the qubit-resonator coupling. The resultant speedup also includes superextensive gates between symmetric Dicke states, characteristic of QB systems. Using a QB eliminates the need for individual drive lines, significantly reducing wiring overhead and potentially quadrupling the number of qubits that can be integrated within cryogenic systems, thereby offering a scalable architecture for quantum computing.

## I. INTRODUCTION

Quantum batteries (QBs) are defined as $d$-dimensional systems that store energy in their excited states while maintaining quantum coherence between the battery levels and the quantum states of their load or charger [1,2]. Over the past decade, this concept has motivated extensive theoretical studies [2], elucidating key battery properties such as enhanced charging power [3,4], bounds on capacity and energy storage [5,6], and extractable work [7]. In parallel, QBs have emerged as a platform to explore collective behavior in solid-state models [8–10] and fundamental aspects of quantum thermodynamics [11,12].

So far, implementations of QBs have focused on demonstrating their superextensive charging property [3,4], where the time reversal of Dicke superradiance [13], i.e., superabsorption, scales the charging (or discharging) power per excitation as $\sqrt{N}$, with $N$ being the number of battery or load quanta. This concept has been experimentally demonstrated in molecules within optical cavities [14,15], superconducting circuits [16–18], and NMR spin systems [19], and has been theoretically proposed for cavity modes [20,21] and many-body systems [10].

Despite the growing interest in QBs as intrinsic energy sources for quantum technologies [22], a practical platform capable of converting stored quantum energy into quantum operations remains absent. This interest is rooted in first-principle quantum thermodynamics since unitary battery-qubit operations create reversible energy transfer during computation, keeping the circuit's energy cost independent of its depth and thereby bypassing the Landauer bound [23]. However, existing proposals for using QBs as the energy source for computation enforce strict separability between the QB and the computational qubits, which introduces a fundamental trade-off between gate fidelity and the QB's energy requirements and results in a QB state that is impractical even for a single computational qubit [24,25].

*Contact author: james.quach@csiro.au

In this work, we present the first framework for utilizing a QB as an intrinsic energy source for quantum computation. We propose a shared bosonic mode acting as the QB, coupled to the computational qubits (Fig. 1). The battery is precharged with sufficient energy to access the full computational Hilbert space, while energy is recycled throughout the computation within the closed qubit–battery entangled system. In a superconducting circuit implementation using a shared-resonator QB, this approach eliminates individual drive lines per qubit, reduces heat load, and may increase qubit capacity per cryogenic system by up to fourfold. We numerically demonstrate that high-fidelity universal quantum computation remains feasible using circuit examples. Gates are implemented by tuning qubit resonance frequencies through multistep protocols: either via direct energy exchange with the QB when operating near its resonance or via virtual excitations in the dispersive regime. In the latter case, we exploit qubit-battery entanglement to realize multiqubit parity probing with a single entangling gate, while the former enables multiqubit unitary gates on shorter timescales. Altogether, this work outlines a concrete route to integrate QBs into quantum processors, creating a new pathway towards energetically efficient and scalable quantum computing.

## II. FRAMEWORK FOR ENERGY-RECYCLING QUANTUM COMPUTATION

### A. Tavis-Cummings Hamiltonian for quantum computation

To investigate how a QB can facilitate quantum information processing, we adapt the well-known Tavis-Cummings model [26] and develop it to reach dressed operators that will govern the quantum logic. The system consists of a bosonic mode acting as the QB, with frequency $\omega_b$ (denoted $\omega_{\text{battery}}$ in the figures), coupled to $N$ two-level systems (qubits) with individual frequencies $\omega_i$. Under the rotating wave approximation, the system is described by the Tavis-Cummings Hamiltonian

$$\hat{H} = \hbar\omega_b \hat{a}^\dagger \hat{a} + \sum_{i=1}^{N} \hbar\omega_i \hat{\sigma}_i^+ \hat{\sigma}_i^- + \sum_{i=1}^{N} \hbar g_i (\hat{\sigma}_i^+ \hat{a} + \hat{a}^\dagger \hat{\sigma}_i^-), \quad (1)$$

with $\hat{a}^\dagger$ and $\hat{a}$ being the bosonic mode's creation and annihilation operators, respectively, and $g_i$ is the coupling constant between qubit $i$ and the battery. The operators $\hat{\sigma}_i^+ = |1\rangle_i\langle 0|_i$ and $\hat{\sigma}_i^- = |0\rangle_i\langle 1|_i$ correspond to the raising and lowering operators of the $i$th qubit. The direct product of the qubit states and the boson number state, $|\vec{s}\rangle \otimes |n_b\rangle$, forms a basis of the quantum computation where $n_b$ represents the number of quanta in the QB and $|\vec{s}\rangle = |s_0 s_1 \ldots s_n\rangle$ are the spin projection quantum numbers of the qubits with $s_i \in \{0, 1\}$.

A key property of the Tavis-Cummings Hamiltonian is its conservation of the total excitation number operator. Because any creation (or annihilation) of a qubit excitation is accompanied by the annihilation (or creation) of quanta in the battery, $[\hat{H}, \hat{n}_{\text{ex}}] = 0$, where $\hat{n}_{\text{ex}} = \hat{a}^\dagger \hat{a} + \sum_i \hat{\sigma}_i^+ \hat{\sigma}_i^-$ counts the total number of excitations in the combined

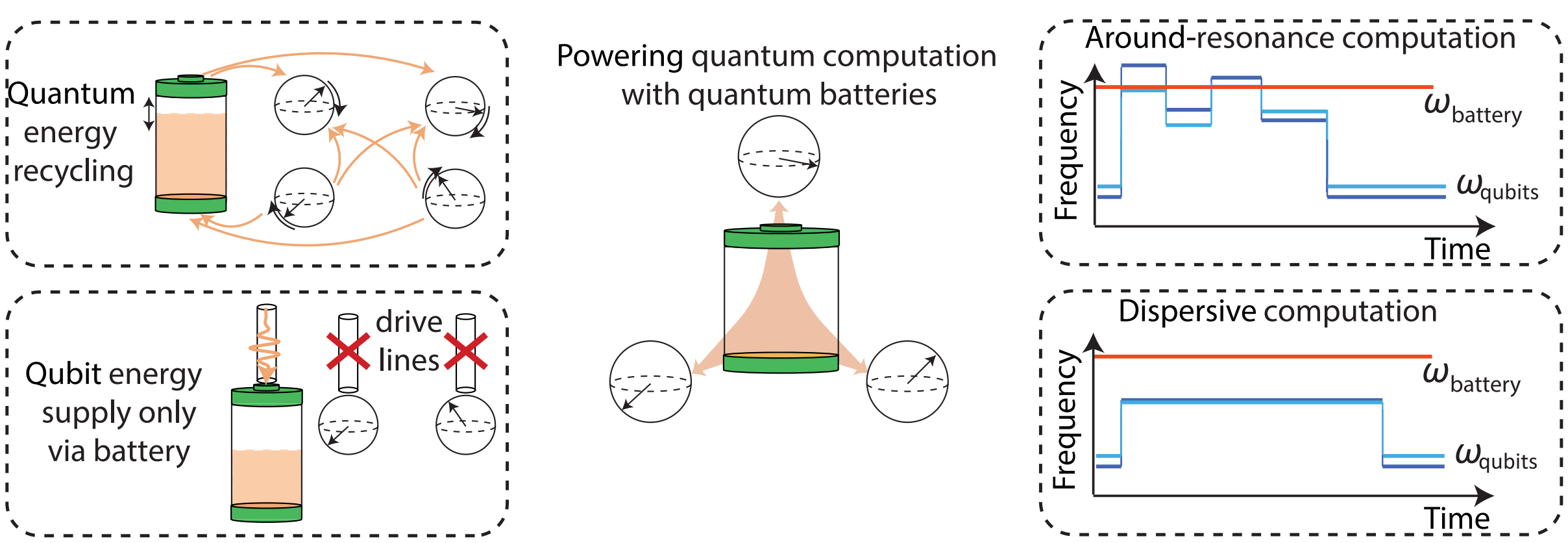

FIG. 1. Powering quantum computing with a QB. A bosonic mode serves as a shared QB for all computational qubits, represented by Bloch spheres (center). The battery is precharged with sufficient energy to implement unitary quantum logic, eliminating the need for individual drive lines (bottom left). This energy is coherently recycled throughout the computation, remaining within the closed qubit-battery quantum system (top left). Quantum computation is implemented by dynamically tuning the qubit resonance frequencies. The QB can facilitate computation both near its resonance (top right) for rapid energy exchange and within the dispersive regime using virtual excitations (bottom right).

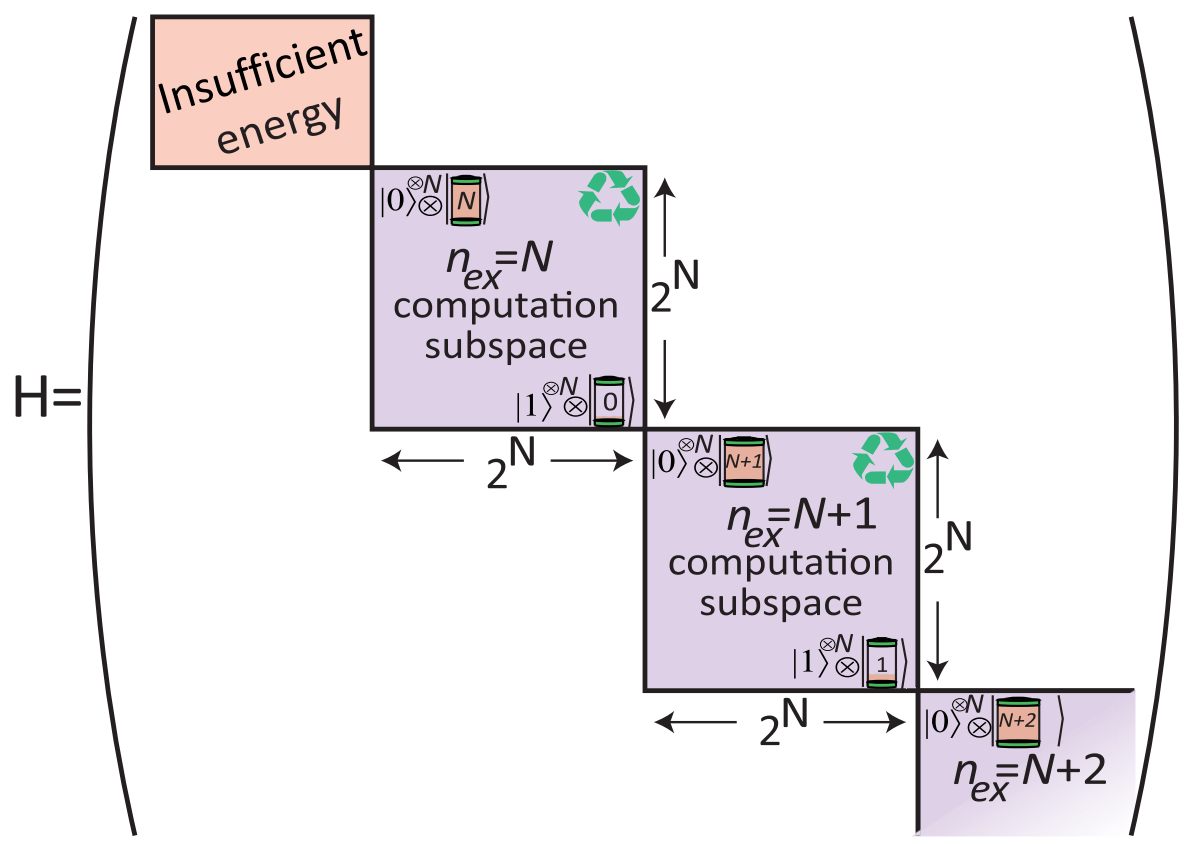

FIG. 2. Block structure of the computational subspaces. Choosing a single value for the total number of excitations in the joint QB-qubits system, $n_{\mathrm{ex}}$, is sufficient to reach a computational subspace, where energy is recycled throughout the computation. When $n_{\mathrm{ex}}$ is smaller than the number of qubits, $N$, the quantum battery lacks sufficient energy to support full quantum computation. Transitions between blocks are only possible through energy loss, injection of external energy, or via counterrotating terms.

QB-qubit system. As a result, the Hamiltonian is block diagonal according to the $\hat{n}_{\mathrm{ex}}$ value (Fig. 2). Assuming the system is initialized in the state $|0\rangle^{\otimes N} \otimes |n_{\mathrm{ex}}\rangle$, where all qubits are in their ground state and the QB is initialized to a bosonic Fock state with $n_{\mathrm{ex}} \geq N$, the time evolution is confined to a single $|n_{\mathrm{ex}}\rangle$ subspace. This subspace has dimension $2^N$, and its dynamics map exactly onto the $2^N$-dimensional Hilbert space of $N$ qubits, the conventional arena for quantum computation, thereby enabling quantum computation with energy recycled entirely within the subspace. We note that, although counterrotating terms can couple between subspaces, their contribution is negligible and can be compensated when $g_i/\omega_i \lesssim 1\%$ for all qubits (derived in Appendix A).

Performing computation within a single excitation-preserving block of the joint system and its energy source introduces new physics for energetically efficient quantum processing. In conventional architectures, the drive is modeled as a classical resource operating far above the single-photon level (including in shared-resonator architectures [27,28]), with multiple attenuators that dissipate substantial heat before the signal reaches the qubits. In contrast, the single-block framework necessarily couples the QB to the qubits, enforcing entanglement between them. Although such entanglement introduces additional decoherence channels, it also enables computations that surpass lower bounds on the energy cost of quantum gates [29]. These bounds were originally derived for classical control fields [30] and later extended to quantum drives [24,25], which require separability between the energy source and the qubits. As we demonstrate below, our approach achieves high-fidelity universal quantum computation even when the QB contains fewer than twice as many quanta as there are computational qubits.

### B. Expressing intrasubspace computation with dressed operators

To formalize the system's evolution during computation within a single excitation-preserving subspace of the full Hamiltonian in Eq. (1), we derive a dressed-operator Hamiltonian. In this representation, the dynamics are expressed entirely in terms of dressed-qubit operators, while the QB degrees of freedom remain implicit. This formulation reveals how the hidden role of the QB generates effective nonlinear terms between qubit operators.

We introduce the QB annihilation operator, $\hat{\sigma}_b^- = (\hat{\sigma}_b^+)^\dagger = \sum_{n_b} |n_b\rangle\langle n_b + 1|$, which satisfies

$$\hat{a}^\dagger |n_b\rangle = \sqrt{n_b + 1}|n_b + 1\rangle = \sqrt{\hat{n}_b}|n_b + 1\rangle = \sqrt{\hat{n}_b}\hat{\sigma}_b^+|n_b\rangle, \tag{2}$$

where $\hat{n}_b = \hat{a}^\dagger\hat{a}$. Equation (2) uses the theorem that if $|n_b + 1\rangle$ is an eigenstate of $\hat{n}_b$ with eigenvalue $n_b + 1$, then it is also an eigenstate of $\sqrt{\hat{n}_b}$ with eigenvalue $\sqrt{n_b + 1}$ [31]. Then, the interacting terms in the Tavis-Cummings (TC) Hamiltonian obey

$$\hat{a}^\dagger\hat{\sigma}_i^-|\vec{s}, n_b\rangle = \sqrt{\hat{n}_b}\hat{\sigma}_i^-|\vec{s}, n_b\rangle = \sqrt{\hat{n}_{\mathrm{ex}} - \hat{n}_q}\hat{\sigma}_b^+\hat{\sigma}_i^-|\vec{s}, n_b\rangle, \tag{3}$$

and similarly, by taking the Hermitian conjugate, $\hat{\sigma}_i^+\hat{a} = \hat{\sigma}_i^+\hat{\sigma}_b^-\sqrt{\hat{n}_{\mathrm{ex}} - \hat{n}_q}$, where $(\sqrt{\hat{n}_{\mathrm{ex}} - \hat{n}_q})^\dagger = \sqrt{\hat{n}_{\mathrm{ex}} - \hat{n}_q}$ is diagonal in the $|\vec{s}\rangle \otimes |n_b\rangle$ basis.

Next, we substitute these interaction terms into the TC Hamiltonian, as well as $\hat{a}^\dagger\hat{a} = \hat{n}_{\mathrm{ex}} - \hat{n}_q$ and $\omega_i = \Delta_i + \omega_b$, where $\Delta_i = \omega_i - \omega_b$ represents the frequency detuning between qubit $i$ and the battery. Then, the system's Hamiltonian becomes

$$\hat{H} = \hbar\omega_b\hat{n}_{\mathrm{ex}} + \sum_{i=1}^{N} \hbar\Delta_i\hat{\sigma}_i^+\hat{\sigma}_i^- + \sum_{i=1}^{N} \hbar g_i\left(\hat{\sigma}_b^-\hat{\sigma}_i^+\sqrt{\hat{n}_{\mathrm{ex}} - \hat{n}_q} + \sqrt{\hat{n}_{\mathrm{ex}} - \hat{n}_q}\hat{\sigma}_b^+\hat{\sigma}_i^-\right), \tag{4}$$

showing how all qubits interact with each other in a nonlinear manner through the QB.

To keep the calculation within a single block of the Hamiltonian and create a direct mapping between the qubits' state and the number of battery excitations, we assume an initial state as $|0\rangle^{\otimes N} \otimes |n_{\mathrm{ex}}\rangle$, corresponding to the ground state of the qubit system and a Fock state with $n_{\rm ex}$ excitations in the resonator. Under this assumption, from now on, we consider $\hat{n}_{\rm ex}$ as a scalar $n_{\rm ex}$.

Within a specific $n_{\rm ex}$ subspace, the system Hamiltonian can be expressed exclusively in terms of dressed-qubit operators, $\hat{\sigma}^+_{d,i} = \hat{\sigma}^-_b\hat{\sigma}^+_i$ and $\hat{\sigma}^-_{d,i} = \hat{\sigma}^+_b\hat{\sigma}^-_i$:

$$\hat{H}_{n_{\rm ex}} = \sum_{i=1}^{N} \hbar\Delta_i \hat{\sigma}^+_{d,i}\hat{\sigma}^-_{d,i} + \sum_{i=1}^{N} \hbar g_i \left(\hat{\sigma}^+_{d,i}\sqrt{n_{\rm ex} - \hat{n}_q} + \sqrt{n_{\rm ex} - \hat{n}_q}\hat{\sigma}^-_{d,i}\right), \quad (5)$$

where we removed the first term as a global phase to all states within the subsystem. In addition, using $\hat{\sigma}^-_b\hat{\sigma}^+_b = \mathbb{I}_{\rm b}$ (the QB identity), the local operator of qubit $i$ can be expressed as $\hat{\sigma}^+_{d,i}\hat{\sigma}^-_{d,i}$ since

$$\hat{\sigma}^+_i\hat{\sigma}^-_i = \hat{\sigma}^+_i\hat{\sigma}^-_i\hat{\sigma}^-_b\hat{\sigma}^+_b = \hat{\sigma}^-_b\hat{\sigma}^+_i\hat{\sigma}^-_i\hat{\sigma}^+_b = \hat{\sigma}^+_{d,i}\hat{\sigma}^-_{d,i},$$

while $\hat{n}_q = \sum_i \hat{\sigma}^+_i\hat{\sigma}^-_i = \sum_i \hat{\sigma}^+_{d,i}\hat{\sigma}^-_{d,i}$ represents the total (dressed-)qubit total excitation operator.

Finally, we show that $\hat{\sigma}^+_{d,i}$ and $\hat{\sigma}^-_{d,i}$ satisfy the Pauli algebra if $n_{\rm ex} \geq N$, i.e., if the battery has enough quanta to populate all qubits to their excited state. To show this, we calculate the commutation relations,

$$\begin{aligned}[\hat{\sigma}^+_{d,i}, \hat{\sigma}^-_{d,i}] &= [\hat{\sigma}^-_b\hat{\sigma}^+_i, \hat{\sigma}^+_b\hat{\sigma}^-_i] \\ &= [\hat{\sigma}^-_b, \hat{\sigma}^+_b]\hat{\sigma}^-_i\hat{\sigma}^+_i + \hat{\sigma}^-_b\hat{\sigma}^+_b[\hat{\sigma}^+_i, \hat{\sigma}^-_i] \\ &= |0_i, 0_b\rangle\langle 0_i, 0_b| + [\hat{\sigma}^+_i, \hat{\sigma}^-_i], \end{aligned} \quad (6)$$

where $|0_i, 0_b\rangle$ denotes the joint vacuum state of qubit $i$ and the bosonic mode. Therefore, the dressed operators preserve Pauli algebra only if the state $|0_i, 0_b\rangle$ is excluded from the subsystem or, equivalently, if $n_{\rm ex} \geq N$. Under this condition, which is necessary for the system to access all qubit states, the commutation relations of dressed operators equal those of qubit operators. As a result,

$$\hat{\sigma}^x_i = \hat{\sigma}^+_{d,i} + \hat{\sigma}^-_{d,i}, \qquad \hat{\sigma}^y_i = i\left(\hat{\sigma}^+_{d,i} - \hat{\sigma}^-_{d,i}\right),$$
$$\hat{\sigma}^z_i = \hat{\sigma}^-_{d,i}\hat{\sigma}^+_{d,i} - \hat{\sigma}^+_{d,i}\hat{\sigma}^-_{d,i}$$

satisfy all standard Pauli operator identities and commutation relations, and serve as the Pauli gates of the quantum circuit.

All quantum information processing presented in this paper is performed using Eq. (5) by manipulating $\Delta_i$ in a stepwise fashion over time and assuming $g_i = g$ for all qubits. This scheme is readily generalized to a distribution of $g$ values and even to their modulation as in tunable-coupler systems [32]. The interaction terms encompass the computation nonlocality through the inseparability between $\hat{n}_q$ and $\hat{\sigma}^\pm_{d,i}$, indicating that the state of all qubits will determine which gate is implemented given a specific set of $\Delta_i$ values and step duration.

## III. UNIVERSAL COMPUTATION IN THE ENTANGLED BATTERY-QUBITS SYSTEM

In this section, we show that the QB-qubit system supports a universal gate set solely through qubit resonance detuning. First, we show how two qubits can be entangled via the QB through a dispersive interaction. Next, we present numerical evidence that the QB can drive a nontrivial population transfer on a single qubit without disturbing the states of the others and independent of the overall system state. By the Solovay-Kitaev theorem [33], this operation, together with arbitrary $Z$ rotations that are implemented by detuning an individual qubit relative to the rest, suffices to generate a universal gate set.

### A. Entangling qubits via dispersive interaction

Entangling qubits via a shared bosonic mode in its vacuum state has become a standard technique in systems governed by the TC Hamiltonian of Eq. (1). The mechanism relies on detuning the target qubits from the mode's resonance by a frequency $\Delta$ in the dispersive regime, where $\Delta \gg g\sqrt{n_{\rm ex}}$. This approach was first established in trapped ions [34] and has since been demonstrated with superconducting qubits [35,36] and spin qubits [37,38]. For two qubits, the interaction implements either an iSWAP or $\sqrt{\rm iSWAP}$ gate, depending on whether the interaction time is set to $t = \pi/(2g^2/\Delta)$ or $t = \pi/(4g^2/\Delta)$, respectively. Moreover, single dispersive gates have been shown to entangle as many as ten [35] and 20 [36] qubits. Here, we extend this picture by presenting an analytic description showing how populating the QB introduces additional degrees of freedom for the resulting entangling gate.

To derive the collective entangling gates, it is convenient to transform Eq. (5) to a Hamiltonian with collective operators, assuming all involved qubits are detuned by the same $\Delta$. The collective operators $\hat{J}^{x,y,z} = \frac{1}{2}\sum_{i=1}^N \hat{\sigma}^{x,y,z}_{d,i}$ satisfy the angular-momentum algebra, $[\hat{J}^i, \hat{J}^j] = i\epsilon_{ijk}\hat{J}^k$, where $\epsilon_{ijk}$ is the Levi-Civita tensor, and $\hat{J}^\pm = \sum_{i=1}^N \hat{\sigma}^\mp_{d,i} = 2(\hat{J}^x \pm i\hat{J}^y)$. Note that the sign flip between + and − arises because $|0\rangle^{\otimes N} = |J = N/2, m_J = J\rangle$, so $\hat{\sigma}^+_i$ increases excitations but reduces $m_J$. Then, the Hamiltonian becomes

$$\hat{H} = \Delta\hat{J}^z + g(\hat{J}^-\hat{A} + \hat{A}\hat{J}^+), \quad (7)$$

where we used $\hat{n}_q = (N/2) - \hat{J}^z$ and substituted $\hat{A} = \sqrt{(n_{\rm ex} - N/2)\mathbb{I} + \hat{J}^z}$, which satisfies $\hat{A}|J, m_J\rangle = \sqrt{(n_{\rm ex} - N/2) + m_J}|J, m_J\rangle$ and $[\hat{A}, \hat{J}^z] = 0$. Since $\hat{J}^2$ remains constant under Eq. (7), the mixture of $J$ values is the source of the collective entangling gates.

To reach the dispersive gates, we perform the Schrieffer-Wolff transformation. For a Hamiltonian $\hat{H} = \hat{H}_0 + \hat{V}$ with $|\hat{V}| \ll |\hat{H}_0|$, the Hamiltonian is approximated to $\hat{H}' \approx \hat{H}_0 + \frac{1}{2}[\hat{S}, \hat{V}]$, where $\hat{S}$ is required to satisfy $\hat{V} + [\hat{S}, \hat{H}_0] = 0$. The operator

$$\hat{S} = \frac{g}{\Delta}(\hat{A}\hat{J}^{+} - \hat{J}^{-}\hat{A})$$

reaches this condition for the Hamiltonian in Eq. (7), so the effective dispersive Hamiltonian then becomes

$$\hat{H}_{\mathrm{disp}} = \left(\Delta + \frac{2g^2}{\Delta}\left(n_{\mathrm{ex}} - \frac{N}{2}\right)\right)\hat{J}^{z} + \frac{2g^2}{\Delta}(\hat{J}^{z})^2 - \frac{g^2}{\Delta}\hat{J}^{-}\hat{J}^{+}, \tag{8}$$

which is similar in form to previous derivations [39–41]. The Hamiltonian conserves the $\hat{J}^z$ value, which is equivalent to the conservation of the number of qubit excitations. Therefore, the gates take an exchange form according to $\hat{J}^{-}\hat{J}^{+}$, which creates interactions between qubit states with a similar number of qubit excitations. A full exchange of quanta occurs when the interaction is applied for $t_{\mathrm{disp}} = \pi/(2g^2/\Delta)$, regardless of the QB's state.

The unique feature of our system is that not only the ratio $g^2/\Delta^2$ and the gate duration but also $n_{\mathrm{ex}}$ determine the executed gate. The QB's state and the detuning value influence the relative phases of different $\hat{J}^z$ values. For a gate duration of $t_{\mathrm{disp}}$, these relative phases exhibit jumps of $\pi/2$ when the ratio $\Delta^2/g^2$ is an integer (as in ion-based systems [42]) and jumps of $\pi$ according to the parity of $n_{\mathrm{ex}}$. Taking a two-qubit example, the dispersive gate is

$$\begin{aligned} &U_{\mathrm{disp}}(n_{\mathrm{ex}}, \Delta) \\ &= \exp(-i\hat{H}_{\mathrm{disp}}t_{\mathrm{disp}}) \\ &= \begin{pmatrix} (-1)^{n_{\mathrm{ex}}} & 0 & 0 & 0 \\ 0 & 0 & -i^{\Delta^2/g^2} & 0 \\ 0 & -i^{\Delta^2/g^2} & 0 & 0 \\ 0 & 0 & 0 & -(-1)^{n_{\mathrm{ex}}+\Delta^2/g^2} \end{pmatrix}, \end{aligned} \tag{9}$$

which implements an iSWAP gate when $\Delta^2/g^2 \bmod 4 = 3$ and $n_{\mathrm{ex}}$ is even.

### B. Single-qubit energy-changing gates

As discussed above, completing a universal gate set relies on reaching a single, local, energy-changing gate that is not a trivial $|0\rangle \leftrightarrow |1\rangle$ mapping, acting on one qubit without disturbing the states of the other qubits. Such a gate may be a Hadamard, a $\sqrt{X}$, or any nontrivial rotation that modifies the qubit's energy. While these gates are routinely realized with high fidelity in conventional architectures using a dedicated drive channel per qubit, they become challenging in a collective QB-qubit system. In this subsection, we show that a small number of detuning steps is sufficient for implementing a local, nontrivial, energy-changing gate that yields the same unitary operation irrespective of the state of the other qubits.

A key property of these gates is that they generate entanglement between the QB and the target qubit. Without loss of generality, we can consider an initial state in which the target qubit is in its ground state and the QB is prepared in a general superposition, $|0\rangle \otimes \sum_{n_b=n_{\mathrm{ex}}-N+1}^{n_{\mathrm{ex}}} C_{n_b}|n_b\rangle$. After the gate, the joint state evolves to $\sum_{n_b=n_{\mathrm{ex}}-N+1}^{n_{\mathrm{ex}}} C_{n_b}(\cos\theta|0\rangle \otimes |n_b\rangle + \sin\theta e^{i\phi}|1\rangle \otimes |n_b - 1\rangle)$, where $\theta \neq 0, \pi$ is the nontrivial energy-transfer rotation angle and $\phi$ is the gate's phase. This entanglement poses an experimental challenge, as QB losses and decoherence are directly mapped onto the qubit state. As we show below, however, allowing such entanglement is essential for achieving lower gate errors per unit of QB energy than the previously established bounds for QB-qubit systems [24].

Such a nontrivial energy-transfer gate can be implemented with perfect fidelity in a two-qubit system coupled to a QB using a single detuning step. To demonstrate this process, we set the target qubit (qubit 0) on resonance with the QB ($\Delta_0 = 0$) while detuning the other qubit (qubit 1) far from resonance ($\Delta_1 \gg g$), such that it acquires only a phase shift without any population transfer. In this configuration, the subsystem formed by qubit 0 and the QB contains $n_{\mathrm{ex}}$ excitations when qubit 1 is in $|0\rangle$ and $n_{\mathrm{ex}} - 1$ excitations when qubit 1 is in $|1\rangle$. The corresponding Rabi frequencies between the QB and qubit 0 are therefore $2g\sqrt{n_{\mathrm{ex}}}$ and $2g\sqrt{n_{\mathrm{ex}} - 1}$, respectively. By choosing the evolution time such that the higher-frequency evolution completes an exact, additional $2\pi$ rotation relative to the lower-frequency evolution, the same gate is applied to qubit 0 irrespective of the state of qubit 1. This condition is satisfied for a minimal gate time $t = \pi/[g(\sqrt{n_{\mathrm{ex}}} - \sqrt{n_{\mathrm{ex}} - 1})]$, which implements a gate of the form $X^{\alpha}$ with $\alpha = \sqrt{n_{\mathrm{ex}} - 1}/(\sqrt{n_{\mathrm{ex}}} - \sqrt{n_{\mathrm{ex}} - 1})$. The nontrivial value of $\alpha$ ensures that this single detuning step suffices to realize a universal single-qubit gate set.

As the number of qubits in the system grows, the dispersion in Rabi frequencies increases, which prevents us from extending the same approach to more than two qubits. For example, in a three-qubit system, there are three distinct Rabi frequencies with irrational ratios between them, making it impossible to realize an exact $2\pi$ rotation in all cases simultaneously. From a theoretical standpoint, any $N$-qubit unitary can be implemented in our system with perfect fidelity, in particular, a single-qubit gate while idling the remaining qubits, using $M \geq (4^N - 1)/(N + 1)$ detuning steps. This condition follows from dividing the total number of degrees of freedom of an $N$-qubit unitary by the number of degrees of freedom available in each detuning step, namely, $N + 1$ (corresponding to $N$ detuning parameters and the step duration). In practice, however, we adopt a numerical approach and show that, even for a five-qubit system, two detuning steps acting only on the target-qubit frequency $\Delta_0$ are sufficient to realize a universal gate set with high fidelity.

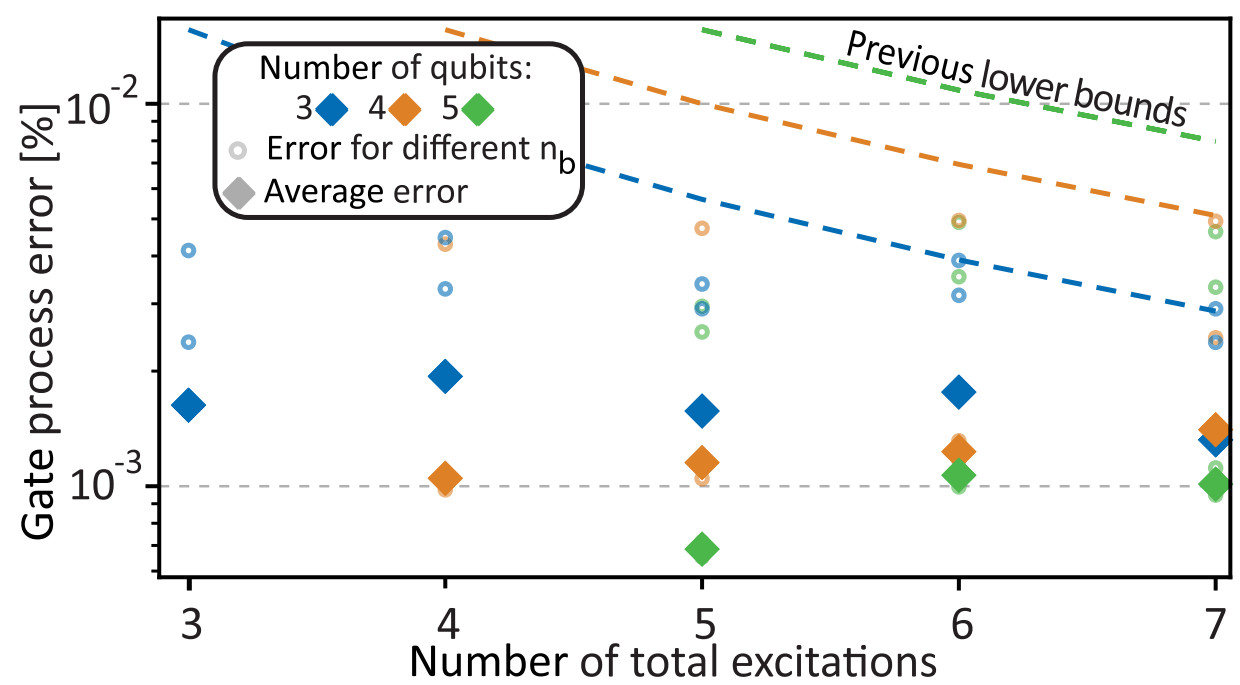

FIG. 3. Gate process error of non-Pauli energy-exchange gates as a function of the total excitation number for multiqubit systems under a two-step detuning protocol. For each combination of qubit number and excitation number, we identify pairs of detuning values and step durations that realize the same energy-changing gate (up to a process error), independent of the initial state of the QB and the remaining qubits. Circles indicate the process error for different initial QB-qubit excitations, while diamonds denote the average error over all such states. All errors lie well below the previously established lower bound from Ref. [24] (dashed lines) because we allow entanglement between the QB and the qubits after the gate ends. Notably, in the two-qubit case, a single detuning step suffices to achieve perfect (100%) fidelity.

Figure 3 shows the resulting process errors for non-Pauli single-qubit gates across different system parameters. For each configuration, defined by the number of qubits and the total number of excitations in the combined QB-qubit system, we numerically optimize the two detuning values and their durations to minimize the gate process error. The diamond markers indicate the average process error for each configuration, while the circle markers show the errors corresponding to different excitation numbers in the QB–target-qubit subsystem. Across all cases shown in Fig. 3, the two-step protocol achieves average fidelities exceeding 99.8% and worst-case fidelities above 99.5%. Notably, each configuration implements a distinct gate.

As an example, consider a three-qubit system with a quantum battery initialized to $n_{ex} = 5$ (blue, center column). We find that applying detunings of $11.58g$ and $-12.32g$ to the target qubit for durations of 38.57 and 40.58 (in units of $1/g$) implements a $0.98\pi$ rotation about the axis $(0.015, -0.36, 1)$ with perfect (100%) fidelity when the other two qubits occupy the $|01\rangle, |10\rangle$ subspace. When the other qubits are instead in $|00\rangle$ or $|11\rangle$, the corresponding process fidelities, evaluated as $F(U_1, U_2) = \frac{1}{4}|\mathrm{Tr}(U_1^\dagger U_2)|^2$ relative to the $\{|01\rangle, |10\rangle\}$ subspace), are 99.82% and 99.77%, respectively, giving an average fidelity of 99.89%. A similar analysis was performed for each configuration of the qubit number and $n_{ex}$, yielding, in every case, a distinct rotation ($\neq \pi$) about a noncardinal axis of the Bloch sphere, implemented using a unique set of optimized detuning values.

Importantly, all errors were found to lie below previous minimal bounds of a single-qubit gate fidelity given the QB input energy. In conventional quantum computing platforms that reach very-low single-qubit infidelities, the qubit drive is deliberately made highly energetic, such that any entanglement generated through energy exchange is negligible compared to the pulse energy, resulting in an almost unmodified scattered drive [43]. In contrast, previous QB studies derived a bound for the minimal QB energy to reach a specific gate infidelity, given by $[N/(8n_{ex})]^2$ (dashed line in Fig. 3) [24], under the strict requirement that the QB and the qubit be separable after the gate. In our approach, this separability constraint is relaxed, and the resulting qubit-QB entanglement enables us to surpass these energy-fidelity bounds, achieving lower gate errors for a given initial QB energy.

Although the errors presented here overcome previously established bounds, they do not represent the fundamental minimum achievable errors. Increasing the number of detuning steps can further improve single-qubit fidelities and, as discussed above, it can, in principle, reach a perfect fidelity based on a degree-of-freedom argument. In experimental setups, lower errors could be achieved through calibration methods such as reinforcement learning. In such a setting, variations in the coupling strength between individual qubits and the QB would likely require separately optimized single-qubit gates for each qubit in each configuration, suggesting that the QB-qubit architecture may be more naturally suited for implementing general $N$-qubit unitaries rather than single-qubit gates. Moreover, in the specific context of optimizing single-qubit operations, a Fock-state QB is not necessarily the optimal initial resource. Alternative initial states, such as coherent states (which require less energy to prepare), may offer advantages, although most of the results presented in this work, which focus on a Fock-state QB, would not directly carry over to that regime.

## IV. EXAMPLES OF QUANTUM CIRCUITS WITH A QUANTUM ENERGY SOURCE

In this section, we present three quantum circuit examples that highlight the advantages of using a QB for quantum information processing. In Sec. IV A, we show how entanglement between the QB and the qubits enables multiqubit control-parity gates in the dispersive regime. Section IV B demonstrates how the combination of all-to-all connectivity and detuning control allows the implementation of multiqubit unitary gates with overall durations of only a few $\pi/g$. Finally, Sec. IV C presents the hallmark features of QBs, such as superextensivity, in specific quantum logic. The specific simulation details are shown in Appendix C.

### A. Dispersive multiqubit parity probing

Parity checks between qubits form the building blocks of most quantum error-correction algorithms, namely, stabilizer codes. Conventionally, probing each parity check between $N$ qubits requires at least $N$ entangling gates. In the circuit example below, we show how using a QB as a quantum energy source enables parity probing with only a single entangling gate, independent of the QB cavity state. This result is possible due to the entanglement between the QB and the qubits, which creates a phase kickback determined by the qubits' parity.

The protocol involves $N$ target qubits, whose parity we wish to probe, and an ancillary qubit, all interacting with the QB with an arbitrary $n_{\mathrm{ex}}$. The sequence begins by entangling the ancillary qubit with the QB using a $\pi/2$ energy-transfer gate. As a result, the joint system of the QB and target qubits is placed in a superposition of subspaces with $n_{\mathrm{ex}}$ and $n_{\mathrm{ex}}-1$, corresponding to the $|0\rangle$ and $|1\rangle$ states of the ancillary qubit. Next, detuning the ancillary qubit and applying a dispersive interaction [Eq. (7)] to the target qubits results in a superposition of $U_{\mathrm{disp}}(n_{\mathrm{ex}})$ and $U_{\mathrm{disp}}(n_{\mathrm{ex}}-1)$, acting on the target qubits, conditioned by the ancillary qubit's state. This approach is equivalent to a controlled operation, where the target unitary

$$U_{\mathrm{ent}}(n_{\mathrm{ex}})^{-1}U_{\mathrm{ent}}(n_{\mathrm{ex}}-1) = e^{2it\frac{g^2}{\Delta}\hat{J}^z} \tag{10}$$

is applied if the ancillary qubit is in the state $|1\rangle$.

By applying the dispersive interaction for a duration $t_{\mathrm{disp}} = \pi/(2g^2/\Delta)$, this target gate becomes $e^{i\hat{J}^z\pi} = (-1)^{\hat{J}^z} = i^N Z^{\otimes N}$. Thus, because of the entanglement between the QB and the ancillary qubit, a controlled-$Z^{\otimes N}$ operation is realized with a single entangling gate (up to a global phase). This controlled-parity gate induces a phase kickback on the ancillary qubit depending on the parity of the target qubits. A subsequent $\pi/2$ gate between the ancillary qubit and the QB then establishes a direct mapping between the $N$-qubit parity and the state of the ancillary qubit. If needed, another dispersive gate can be used to restore the target qubits to their original state. Overall, this protocol enables the probing of any $Z^{\otimes N}$ parity operator with a single dispersive gate, independent of the QB's state and the number of qubits. Including local gates, we can extend this probing to $X^{\otimes N}$, completing the set of stabilizers required for QEC codes.

Figure 4 demonstrates a specific circuit (a) and its simulated implementation (b) when probing two-qubit parity. In this simulation, we choose an $I\otimes I\otimes H$ gate to create an entanglement between the ancillary qubit ($q_2$) and the QB and then disentangle them after the controlled-parity gate. The dispersive gate implements CZ and SWAP gates, which are accompanied by a set of local $Z$ gates if qubit $q_2$ is in the $|1\rangle$ state. Eight detuning steps were used to implement the $I\otimes I\otimes H$ gate, with an overall duration of $\pi/4g$ [inset of Fig. 4(b)]. In the dispersive gate (and the restoring step), $q_{0,1}$ were detuned to $\Delta = 27.5g$ for a duration of $27.5\pi/2g$, while $q_2$ was heavily detuned to minimize its effect. Comparing the simulated circuit with an ideal parity-probing circuit [Fig. 4(a)], we obtain a Hilbert-Schmidt distance of 0.642, corresponding to an average circuit fidelity of 95.4%. Fidelity could be further improved by allowing continuous time evolution, which adds degrees of freedom.

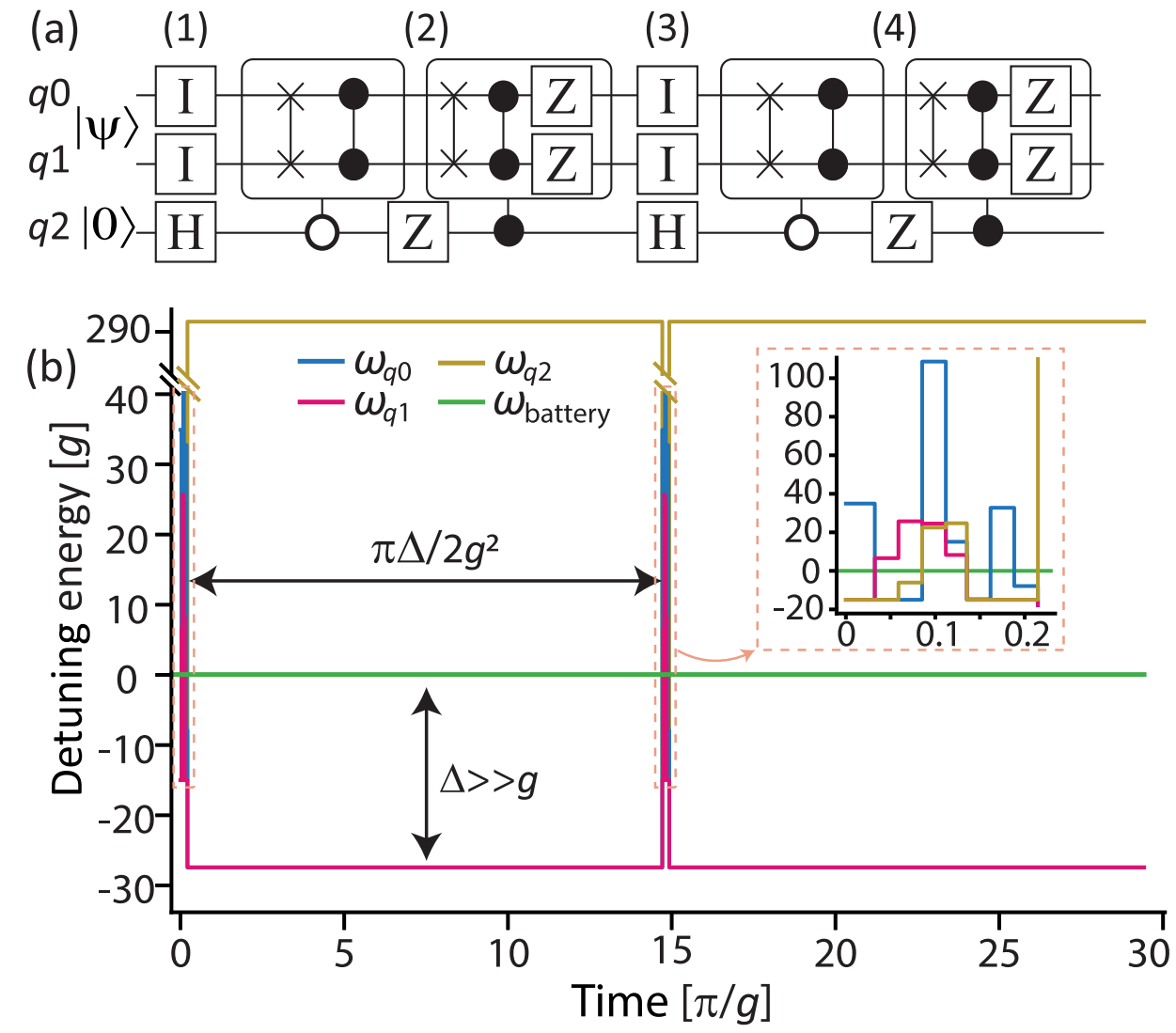

FIG. 4. Multiqubit parity probing through battery-dependent dispersive gates. (a) Protocol for probing multiqubit parities through dispersive interactions, exemplified with a two-qubit probing circuit. Step (1) entangles the ancillary qubit ($q2$) with the battery. Step (2) applies a battery-dependent dispersive gate that implements a superposition of $U_{\mathrm{disp}}(n_{\mathrm{ex}})$ and $U_{\mathrm{disp}}(n_{\mathrm{ex}}-1)$ in the joint battery–two-qubit subspace. Step (3) maps the parity of $|\psi\rangle$ onto the ancillary qubit, producing $|\psi\rangle_{\mathrm{even}}|1\rangle_{q2} + |\psi\rangle_{\mathrm{odd}}|0\rangle_{q2}$, where $|\psi\rangle_{\mathrm{even}}$ and $|\psi\rangle_{\mathrm{odd}}$ are the even and odd components of $|\psi\rangle$, respectively. Step (4) restores the probed qubits to their original state $|\psi\rangle$. This sequence remains unchanged for any $N$-qubit parity probing. (b) Implementation of the circuit in panel (a) using only qubit frequency detunings, for $n_{\mathrm{ex}}=5$. The duration of each dispersive gate is $(\Delta/g)[\pi/(2g)]$ (with $\Delta = 27.5g$), independent of the number of qubits probed. The $I\otimes I\otimes H$ gate takes approximately $\frac{1}{2}[\pi/(2g)]$, implemented here in eight steps (inset).

### B. Around-resonance multiqubit gates

In the previous subsection, we have shown how the entanglement between a populated QB and the qubits can be utilized for unique processing. However, the timescales of the dispersive gates might introduce errors in their experimental implementation, especially when increasing the number of qubits [36]. This effect motivated efforts to find new computation schemes for shared-resonator quantum computers using driven cavities [28]. In this section, we show that the QB can facilitate computation without reaching the dispersive regime timescales but rather through energy exchange processes between the QB and the qubits, reaching overall unitary durations of a few $\pi/g$.

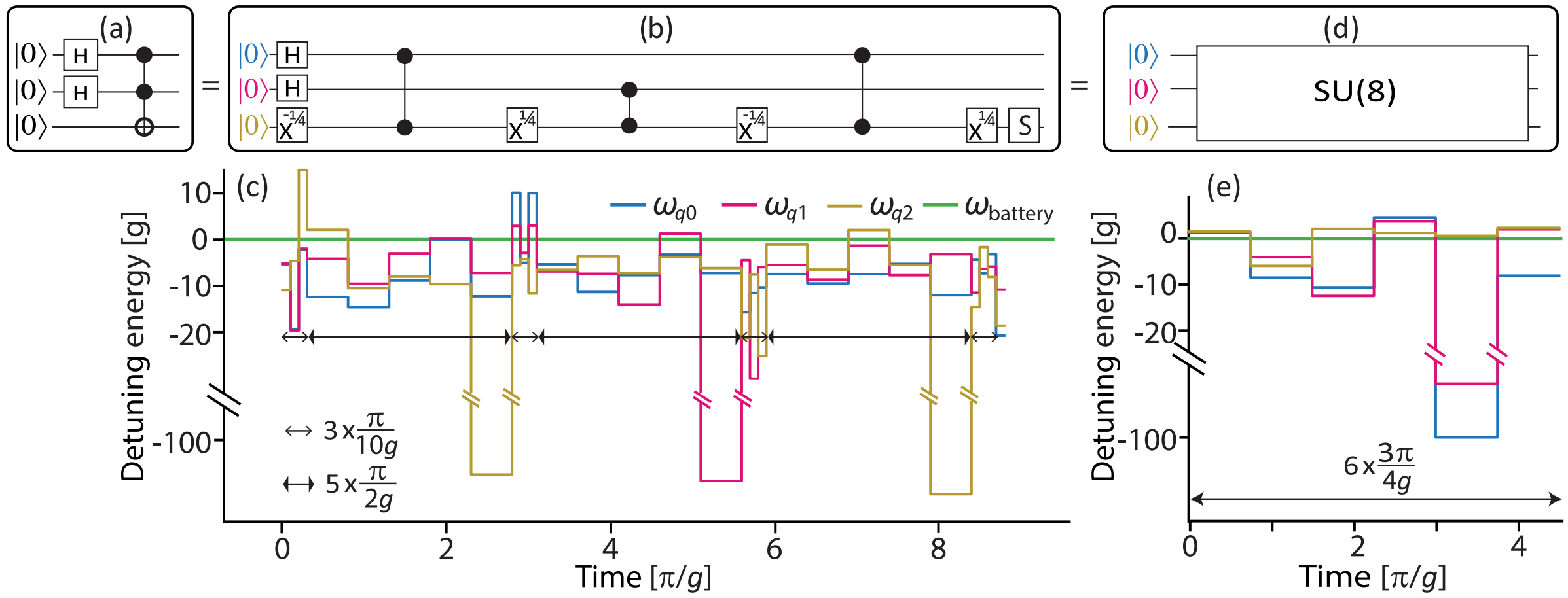

FIG. 5. Computation via battery-mediated multiqubit interactions near resonance. (a) Target circuit featuring a three-qubit Toffoli gate, applied after preparing the control qubits in superposition. (b) Standard decomposition of the target circuit into a sequence of single-qubit and pairwise entangling gates. (c) Implementation of panel (b), yielding a final infidelity of $3 \times 10^{-3}$. Energy-changing operations (thin arrows) are realized in three detuning steps, each with a duration of $\pi/10g$, while entangling operations (thick arrows) require five steps, each of duration $\pi/2g$, to account for round-trip energy exchange via the battery. (d) Uniting the full target circuit into a single multiqubit unitary. (e) Implementation of panel (d) using six steps of duration $3\pi/4g$, reaching an infidelity of $3 \times 10^{-9}$ while significantly reducing the total circuit time compared to panel (c).

Figure 5(a) presents a circuit example that involves energy transfer from the QB to the qubits and a non-Clifford three-qubit entangling gate in the form of a Toffoli gate. In conventional quantum processors, this circuit is broken down to the hardware's native gates as shown in Fig. 5(b). We start with a similar approach to implement the circuit using an interleaved sequence of energy-transfer gates and entangling CZ gates, shown in Fig. 5(c). We implement each energy-transfer gate in three $0.1\pi/g$ steps. The CZ gates are implemented through five steps of $\pi/2g$, intuitively to enable a back-and-forth interaction between two qubits via the QB. Overall, this simulation implements the circuit with an overall infidelity of $3 \times 10^{-3}$, where the detuning parameters are chosen to optimize the fidelity.

Figure 5(e) shows that the overall circuit time can be further reduced and the implementation can be simplified when combining the whole circuit into a single SU(8) unitary [Fig. 5(d)] implemented in a few detuning steps. Specifically, we use six $3\pi/4g$ steps to reach an infidelity of $3 \times 10^{-9}$ with an overall duration of $4.5\pi/g$. This example shows that the shared resonator can combine multiqubit entangling and energy transfer from the QB to the qubits within the same sequence. This property can open paths to new computation methods with smaller circuit depth that utilize the all-to-all connectivity of the shared-resonator architecture, as opposed to a shared-resonator pairwise computation [28].

### C. Quantum battery features during computation

This subsection focuses on how the system supports the hallmark characteristic of QBs—mainly, their superextensive speedup in energy transfer [3,4,14] and the improved fidelity when increasing the QB population. We analyze the implication of this property on quantum computation when setting $\Delta_i = 0$ for all qubits, reducing the Hamiltonian to $\hbar g \sum_{i=1}^{N} (\hat{\sigma}_{d,i}^{+}\sqrt{n_{\text{ex}} - \hat{n}_q} + \sqrt{n_{\text{ex}} - \hat{n}_q}\hat{\sigma}_{d,i}^{-})$. In the single-qubit case, this Hamiltonian simplifies to $\hbar g \sqrt{n_{\text{ex}}}\hat{\sigma}_{d,i}^{x}$, recovering the expected Jaynes-Cummings result of executing an $X$ gate after a $t = \pi/(2g\sqrt{n_{\text{ex}}})$ interaction time. This case can be generalized to any QB-qubit interaction where only qubit $i$ is on resonance and the others are highly off-resonance so that an initial state of $|n_b\rangle \otimes |0\rangle_i$ in the QB-qubit subspace reaches $|n_b - 1\rangle \otimes |1\rangle_i$ in $t = \pi/(2g\sqrt{n_b})$ [dashed line in Fig. 6(a)].

The superextensive speedup, shown in Fig. 6(a), arises as qubits become on resonance with the QB, enabling reduced gate times for mapping any two symmetric Dicke states of the system, such as $|0\rangle^{\otimes N} \leftrightarrow |1\rangle^{\otimes N}$. The gate time decreases when increasing the number of qubits, assuming a fixed number of battery quanta per qubit ($n_b/N$) in the initial state $|n_b\rangle \otimes |0\rangle^{\otimes N}$. Increasing this ratio will reduce the gate time, eventually converging to the superextensive speedup limit of $1/\sqrt{N}$. The time reversal of the simulated gate is the famous Dicke superradiance, which was demonstrated in many quantum computation platforms with a vacuum-state shared mode [44,45]. Our scheme shows a use case in which QBs harness the collective effects of superradiance and superabsorption to accelerate quantum computation.

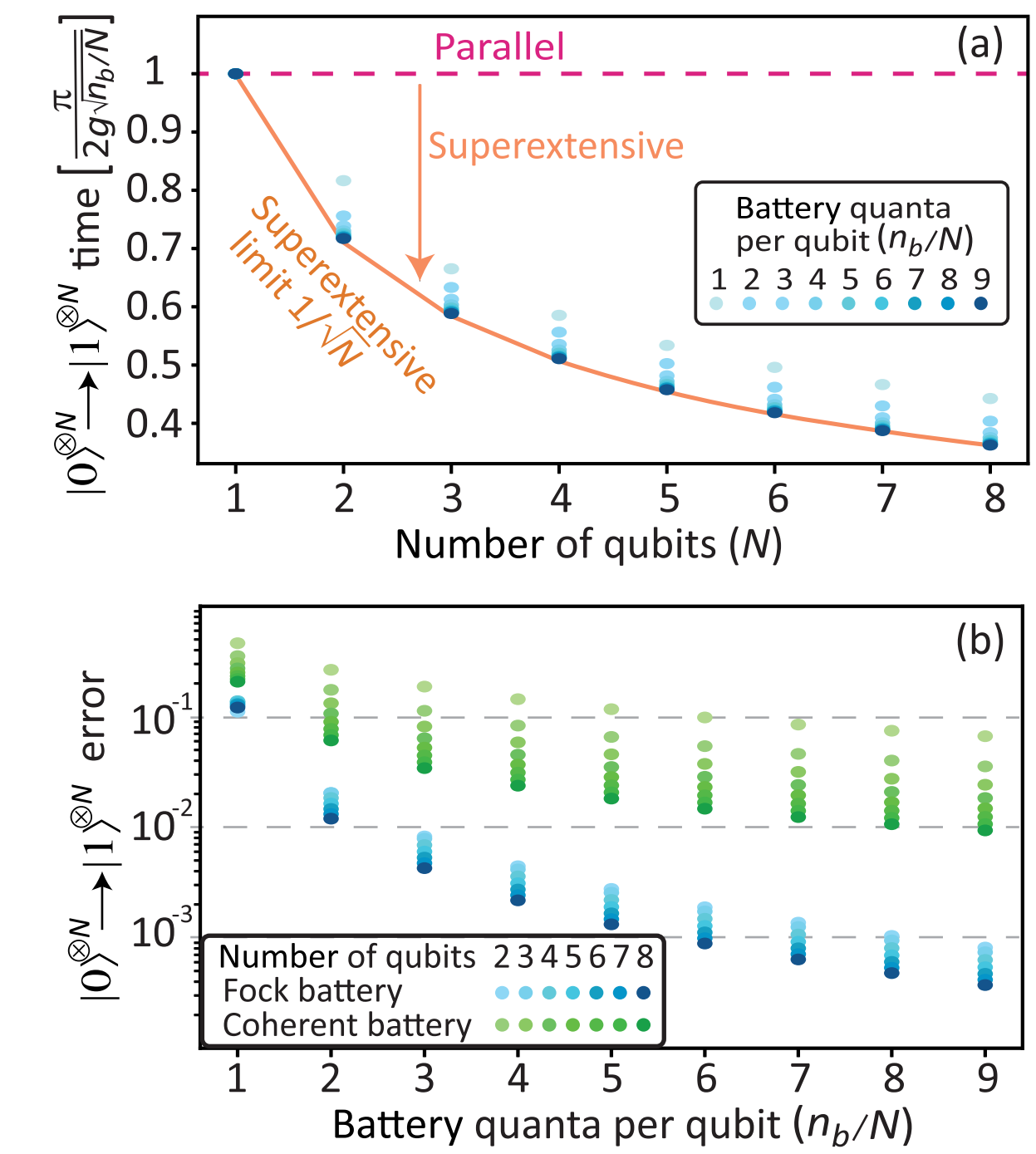

FIG. 6. Superextensive scaling in a $|0\rangle^{\otimes N} \to |1\rangle^{\otimes N}$ gate. (a) Gate time for different battery-quanta-per-qubit ratios (blue dots), normalized to the parallel charging time ($N$ separate batteries, each coupled to a single qubit; pink dashes). Values below 1 indicate superextensive speedup, where collective multiqubit interactions reduce gate times when increasing the number of qubits, given a fixed $n_b/N$ ratio, until reaching the superextensive speedup limit of $1/\sqrt{N}$ (orange). (b) Collective gate error as a function of the initial number of QB quanta when initialized in a Fock state (blue shades) and in a coherent state (green shades). The Fock-state QB exhibits significantly better performance, with the gate error decreasing when adding qubits. In both panels (a) and (b), all qubits are on resonance with the QB.

Another famous feature of QBs comes from the increased precision of these collective gates when increasing the number of qubits [20]. Figure 6(b) shows that, when the QB is initialized in the Fock state, the gate-error values are between $10^{-2}$ and $10^{-3}$ [blue shades in Fig. 6(b)]. Such values are unattainable when the QB is initialized in a coherent state, where the $n_b$ values are coherently distributed in the initial state [green shades in Fig. 6(b)]. These results highlight the importance of the QB being initialized in a quantum state of light [46,47] rather than a coherent state, overcoming previous bounds on the gate fidelities and minimal energy requirements, which assumed an external coherent energy source [48] or a QB that is separable from the qubit system [24].

## V. EXPERIMENTAL FEASIBILITY WITH SUPERCONDUCTING QUBITS

In this section, we discuss the feasibility of experimentally realizing our schemes, focusing on an implementation with superconducting transmon qubits and addressing key considerations.

To assess resilience against system noise, we simulate the detuning sequence from Fig. 5(e) under realistic noisy conditions, achieving a final-state fidelity above 99.2%. The system's evolution is modeled with typical superconducting parameters of $g = 15$ MHz, corresponding to a total evolution time of 150 ns (six 25-ns steps), a cavity lifetime of 600 μs [49], single-qubit $T_1$ and $T_2^*$ times of 30 μs and 10 μs, respectively (conservative compared with state-of-the-art examples [50]), and a five-photon Fock-state purity of 99%. The simulated fidelity remains unchanged even when including additional phase drifts with a standard deviation of 0.005 rad/ns. These results suggest that current superconducting hardware has reached a maturity level sufficient, in principle, to demonstrate quantum information processing with QBs.

High-fidelity initialization of the QB in a Fock state can be achieved with several existing protocols [46,51,52]. Any charging protocol requires a nonlinear element, which can be implemented using one of the qubits coupled to the QB, with its drive line acting as the drive to the QB. Recent proposals demonstrated the preparation of five-photon Fock states with near-perfect fidelity when incorporating real-time feed forward [52]. Maintaining a small Fock number is essential, as the QB lifetime scales as $1/n_b$ [46]. This requirement is especially crucial in our computation scheme, as any QB decoherence is converted to qubit decoherence due to their entanglement during the computation. The circuit examples in this work use $n_b/N < 2$ to achieve low-error unitary operations by optimizing detuning values over multiple steps. Moreover, since our protocols operate on timescales of order $\pi/g$, the additional errors due to photon loss remain manageable.

Executing computations requires calibration of the optimized flux parameters for each device, due to fabrication variances. In particular, flux profiles must avoid ringing or overshoot [53] by smoothing the signals during rise and fall times, which can be on the order of 5 ns to be compatible with the control electronics and the typical timescales of our scheme. Similar flux signals and calibration methods have been demonstrated in shared-resonator qubit architectures [28], achieving high fidelity with the aid of a tunable coupler. Such couplers add an additional flux line per qubit but also provide an extra tunability parameter via the coupling constant, which is widely used in state-of-the-art superconducting processors [54]. In these implementations, gates are realized by jointly tuning the flux of the coupler and the qubit.

Finally, we use the parameters of Fig. 5(e) to evaluate the contribution of counterrotating terms beyond the assumed

Tavis-Cummings Hamiltonian (see Appendix A). The procedure for quantum computation remains quantitatively accurate as long as the coupling ratio satisfies $g/\omega_b \lesssim 1\%$. As shown in Fig. 8, recalibrating the step durations by 0.05% per step can reduce errors from counterrotating terms by nearly 2 orders of magnitude. Their small effect arises because the counterrotating terms do not create energy transfer between blocks since at least 100 counterrotating oscillations average out during a detuning step duration of about $\pi/g$.

## VI. ENERGETIC EFFICIENCY AND SCALING OPPORTUNITIES

Powering quantum computers with QBs offers a unique opportunity to scale cryogenic processors by providing full qubit control through flux lines and supplying all unitary gate energy prior to the computation. This approach relies on fundamental principles, enabling an architecture with fewer hardware components and consuming less energy during computation.

At a fundamental level, powering quantum computers with QBs has the potential to reach the limit of zero active heat generation during unitary logic. The von Neumann entropy of a quantum system remains unchanged under any unitary transformation and is zero for any pure state. Consequently, since entropy does not change during a unitary quantum circuit, computation can, in principle, proceed without heat generation [23]. This case follows from the thermodynamic bound $\Delta Q \geq -T\Delta S$, where $\Delta Q$ is heat generation, $\Delta S$ is entropy change, and $T$ is temperature. Achieving this bound is impossible with conventional external drive pulses due to attenuation losses and the minimal energy supply needed for each quantum operation [24,30]. However, using a closed QB-qubit system and allowing entanglement between the two, one can approach the limit by reusing the energy charge in the QB prior to the circuit. This case highlights the potential of QBs to enable energy-efficient quantum computation and to move towards thermodynamically optimal quantum processing.

To quantify how these fundamental arguments translate into heat dissipation in realistic macroscopic systems, we compare the active energy and heat generated by quantum circuits in conventional and QB-based architectures. The inset of Fig. 7(a) illustrates the key distinction: In the QB approach, qubits couple to a shared resonator rather than to individual drive lines, thereby eliminating both the active and passive heat sources present in conventional schemes. This reduction in interconnect-related heat directly increases the number of qubits that can be supported per cryogenic refrigerator. In this way, the principle of zero-entropy gates with a QB maps onto a concrete chip-level architecture, enabling quantum computations to be executed with reduced energetic cost and alleviating a central scalability bottleneck in cryogenic quantum computing.

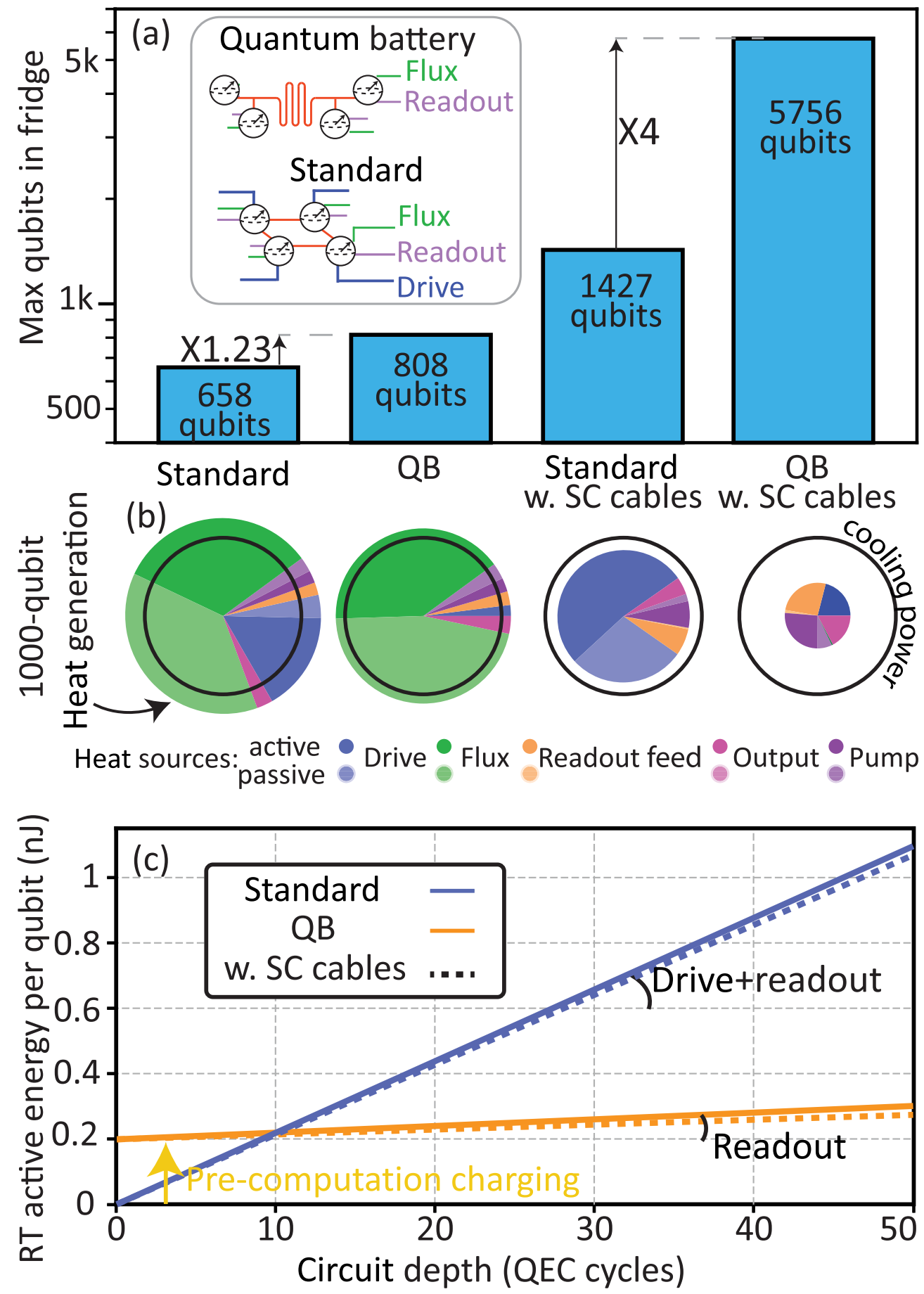

FIG. 7. Scaling opportunities and energy requirements of the QB architecture with superconducting qubits. (a) Maximum number of qubits that the state-of-the-art cryogenic fridge can support, derived from the minimal number of qubits that reach the cooling power. The QB configuration enables a factor of 1.23 additional qubits with available cables, while superconducting flux lines will enable one to increase the number of qubits by a factor of 4 compared to the standard configuration (with superconducting cables). Inset: QB (top) and the conventional (bottom) architectures. The conventional computation includes an additional control line per qubit (drive) and corresponding attenuators that create passive and active heat during the computation. (b) Distribution of heat sources at the cryogenic fridge limiting levels. The area of each pie chart represent the total heat power of a thousand qubits compared to the cooling power (black). (c) Accumulated room-temperature (RT) energy per qubit required for the input control lines as a function of the circuit depth. Although the QB needs to be charged prior to the computation, removing the drive pulses allows us to keep the additional energy to be dominated by readout, and the QB computation becomes energetically efficient after a few quantum error correction (QEC) cycles. A detailed derivation of panels (a) and (b) is provided in Appendix B.

By comparing the total heat load to the state-of-the-art cooling capacity of cryogenic fridges, we assess the potential scaling advantages of the QB architecture, as illustrated in Fig. 7(a). To evaluate how flux-only control affects scalability, we estimate the heat power dissipated at the two coldest stages of a dilution refrigerator during quantum computation, following the methodology of Krinner *et al.* [55] (see Appendix B and Fig. 9 for details). Figure 7(b) displays the total heat generated by a 1000-qubit system (represented by the area of the pie chart), from which the maximum number of qubits per fridge can be inferred by comparing it to the fridge's available cooling power (black).

We calculate the active and passive heat contributions from control lines based on standard pulse usage [colored segments in Fig. 7(b)]. This analysis reveals that, with current cable technology, adopting a QB architecture can increase the number of qubits per fridge by 23%, with the dominant limitation being attenuation along the flux lines at the mixing chamber stage. If superconducting cables are used for the flux lines, the scaling potential improves dramatically: We observe an 8.75-fold increase compared to the standard configuration and a fourfold improvement compared to a conventional setup with superconducting cables. These gains are not achievable if drive lines are retained, as their associated attenuators generate heat, which prevents the use of superconducting materials. In the QB configuration with superconducting cables, heat dissipation is primarily due to the readout components, including feed lines and amplifier pumps, while the remaining heat arises from passive dissipation along the single drive line per battery (assuming each battery controls ten qubits).

In addition to the quantum computer scaling opportunities, QBs reduce the total active energy consumption of the quantum computation. To quantify this process, we calculate the active power that enters the cryogenic fridge to estimate the computation's input energy at room temperature (detailed in Appendix B). The results, presented in Fig. 7(c), show that the QB architecture becomes more energy efficient than conventional designs once the circuit depth exceeds the energy cost of preparing the QB in the Fock state. We conservatively estimate this preparation energy as equivalent to the required energy for driving and a readout of ten qubits over ten QEC cycles, based on the energy needed to overcome competing losses [56].

Once initialized, the QB architecture eliminates drive-pulse energy consumption during computation, leaving only readout pulses as an energy input. Consequently, the energy per circuit depth accumulates more slowly than in conventional architectures, and the energy consumption gap between the two approaches widens as the circuit depth increases. At a fundamental level, QEC protocols use midcircuit readout to stabilize the qubits, so the only energy entering the system during QB-based computation is used solely to remove entropy.

Our heat and energy results can be directly incorporated into previous holistic derivations of the energy cost of quantum computation. While most of the energy expenditure in quantum computers is still expected to be dominated by cryogenic cooling rather than active computation [57,58], the balance between cooling and computation depends strongly on system architecture. The large drive channel-to-qubit multiplexing of the QB architecture meets the expected high-level requirement for energetically efficient quantum computation [59]. In particular, the qubit-enhancement ratios we present for a fixed cooling power can equivalently be interpreted as reduced cooling demand for a given quantum algorithm. From an algorithmic perspective [60,61], our architecture decreases the average power per qubit, which can be further reduced through circuit-level optimization [61].

## VII. DISCUSSION

In this paper, we presented the first practical framework for integrating QBs as intrinsic energy sources for quantum technologies, in particular, for quantum computation. We demonstrate that a precharged shared bosonic mode, which becomes entangled with the qubits during computation, enables universal quantum computation through frequency detuning alone. To the best of our knowledge, this work is the first proposal to harness stored quantum energy in a bosonic mode to perform quantum logic, relying solely on frequency detuning. By recycling energy within a closed QB-qubit system, our scheme approaches the thermodynamic limit where energy is consumed only during readout. In cryogenic platforms, eliminating individual drive lines enables a scalable control architecture built from standard components, offering a practical alternative to more complex scalability proposals such as photonic links [62].

Several challenges remain before large-scale QB-based quantum computing can be realized. To execute computations, we optimized flux parameters using off-the-shelf routines in our simulations; however, scaling this process to larger systems may demand more advanced optimization and calibration methods. As discussed in Sec. V, the proposed QB architecture is compatible with existing quantum hardware and classical control systems where initializing the Fock state and calibrating the gates will benefit from real-time feedback. Importantly, our protocol reaches circuit timescales on the order of $\pi/g$, allowing for fast entangling gates outside the dispersive regime and implementing high-fidelity quantum circuits even when including realistic system noise, which we expect to be suppressed when eliminating drive lines.

Our approach is facilitated by a simplified control and reduced thermal load. Removing a dedicated drive line per qubit at cryogenic temperature eliminates two associated lines per qubit at room temperature, thereby reducing the overall energy consumption, lowering system complexity, and minimizing signal crosstalk [63]. Moreover, traditional analog drive control requires high precision, which needs a significant electronic overhead, in turn forcing the electronics to remain at room temperature. In contrast, frequency-based control aligns naturally with emerging cryogenic-compatible technologies, such as cryo-CMOS [64] or single-flux-quantum logic [65], offering additional reductions in power consumption. In addition, integrating QBs with efficient readout methods, such as microwave photomultipliers [66], can further enhance the scaling potential. Altogether, incorporating QBs into quantum processors represents a promising pathway toward simpler, more scalable, and energetically favorable quantum computation.

## ACKNOWLEDGMENTS

K. H. and J. Q. Q. acknowledge funding from the Revolutionary Energy Storage Systems Future Science Platform. W. J. M. was partially supported by the MEXT Quantum Leap Flagship Program (MEXT Q-LEAP) under Grant No. JPMXS0118069605.

## DATA AVAILABILITY

The data that support the findings of this article are openly available [67]; embargo periods may apply.

## APPENDIX A: CONTRIBUTION OF COUNTERROTATING TERMS

In the derivation in Sec. II, we have restricted our description to the rotating-wave approximation (RWA), where the system Hamiltonian is block diagonal in the excitation number basis and all terms conserve the total number of excitations. Relaxing this approximation introduces additional counterrotating terms in Eq. (1), of the form

$$\hat{H}_{\mathrm{CR}} = \hbar g \sum_i \left(\hat{a}\hat{\sigma}_i^- + \hat{a}^\dagger \hat{\sigma}_i^+\right), \tag{A1}$$

which couple states belonging to different excitation-number sectors. Physically, these terms allow for processes that simultaneously create or annihilate excitations in both the qubit and the resonator, thereby breaking the conservation of the total excitation number. Such processes become relevant when the coupling strength $g$ approaches the characteristic transition frequencies $\omega$ of the system. In this regime, the rotating-wave approximation is no longer strictly valid, and the dynamics can deviate from those predicted by the block-diagonal Tavis-Cummings description.

To quantify their impact in our setting, we performed simulations including the full Hamiltonian with counterrotating terms, shown in Fig. 8. When keeping the detuning parameters exactly as in Fig. 5(e), the infidelity grows systematically with the ratio $g/\omega$, reaching the percent level already at $g = 0.003\omega$ (corresponding to $g = 15$ MHz for a resonator frequency of 5 GHz). Nevertheless, a modest reoptimization of the detuning times, while keeping the detuning amplitudes fixed, is sufficient to suppress the counterrotating errors by nearly 2 orders of magnitude for specific working points of $g/\omega$. This correction only requires small adjustments of the timestep to compensate for relative phase accumulation induced by the counterrotating terms. The effectiveness of such a minimal correction indicates that our scheme is robust against moderate deviations from the RWA regime. However, when the coupling strength exceeds the percent level of the transition frequency, further parameter reoptimizations, which go beyond our simple time adjustments, will be necessary to maintain high-fidelity operation.

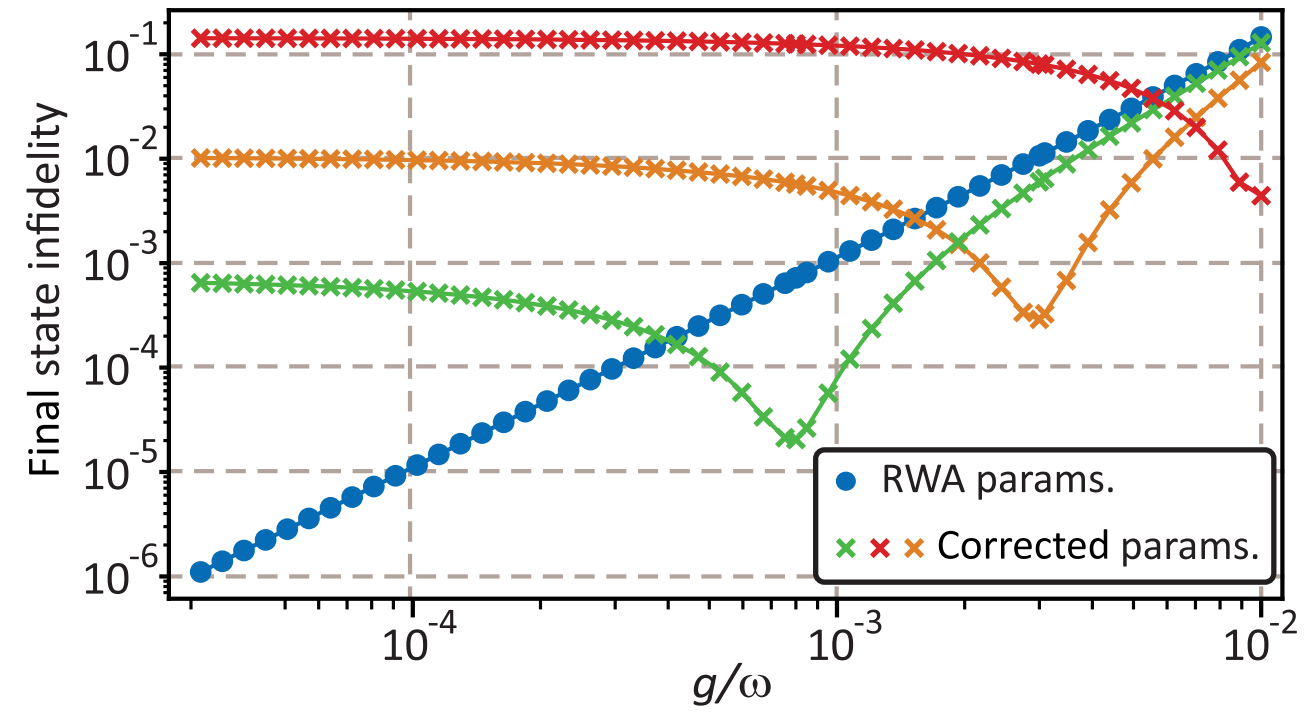

FIG. 8. Effect of counterrotating terms. We show the infidelity of the circuit from Fig. 5 when counterrotating terms are included. The blue curve corresponds to the original detuning parameters in Fig. 5(e), showing, on a log-log scale, how the infidelity increases with the coupling ratio $g/\omega$, reaching an error of around 1% once the ratio reaches $g = 0.003\omega$. The other curves show how a reoptimization step of only the original detuning times for the full non-RWA Hamiltonian can reduce infidelities by almost 2 orders of magnitude, when provided a target $g/\omega$ value (green: $g = 0.0008\omega$; orange: $g = 0.003\omega$; red: $g = 0.01\omega$). The reoptimization step modifies the time step by a small factor, only to overcome the relative phase accumulation due to the counterrotating terms. Further error reduction when $g/\omega > 1\%$ will require a more significant parameter reoptimization.

Overall, from these simulations, it is clear that the counterrotating terms do not prevent quantum computation with quantum batteries. When $g/\omega$ is below 1%, which is expected in typical superconducting circuit hardware (e.g., $g < 50$ MHz), a small correction to parameters derived from the RWA Hamiltonian is sufficient to suppress the contribution of the counterrotating terms. For $g/\omega$ values above 1%, a shared resonator can still provide energy to the computational qubits, although a more significant parameter optimization is required, and the block-diagonal description (including the dressed-operator representation) can no longer be employed.

## APPENDIX B: HEAT DISSIPATION ANALYSIS

The goal of this appendix is to compare the conventional quantum computation to the computation powered by QBs, in terms of the created heating power in the bottom two stages of the cryogenic fridge, based on the analysis from Ref. [55]. We concentrate on the lowest two layers in the cryogenic fridge [cold plate (*CP*) and mixing chamber (MXC)] because they are the hardest to cool and are considered the limiting stages on qubit count. We estimate the qubit count when comparing the generated heat to the cooling power in the state-of-the-art cryogenic system (Bluefors XLD1000sl).

#### 1. Quantum computation channel types

The configuration described here is oriented toward flux-tunable superconducting qubits, though a similar analysis applies to fixed-frequency qubits and semiconductor spin qubits. We use five types of channels: drive, flux, readout resonator feed, readout output, and readout amplifier pump. Drive channels control XY gates with GHz analog modulated pulses sent through coaxial lines capacitively coupled to the qubit; each qubit has a unique drive channel, and these channels include large attenuation (20 dB at both MXC and *CP*), generating significant heat without prospects for reduction. Our QB approach removes these channels. Flux channels carry currents to set qubit frequencies, requiring stable dc or up to 1–2 GHz signals; with superconducting wiring and minimal attenuation, they contribute negligible heat. Readout resonator feed channels apply microwave signals to resonators coupled to qubits, with attenuation-like drive lines but lower signal amplitudes, resulting in a similar amount of passive heat, but less active heat. Readout output channels use superconducting cables without attenuators, contributing negligible heat. Readout amplifier pump channels drive cryogenic amplifiers (e.g., TWPAs) with a signal of around −55 dBm in the MXC; these lines have 10 dB less attenuation in *CP* than drive lines but require significantly more power, making their active load comparable to drive channels.

We consider three cable options: stainless steel coaxial, niobium-titanium, and optimal superconducting. Stainless steel coaxial cables (UT-085-SS-SS) are used for channels with attenuators. Niobium-titanium cables (UT-085-NbTi) are used for channels without attenuators; we use passive loads from theoretical calculations and assume negligible active loads when superconducting. Optimal superconducting cables are analyzed as twisted-pair wires with passive loads estimated from Ref. [55].

In this derivation, we compare the heat generated during quantum computation among the standard architecture, the QB architecture with standard cables, and the QB architecture with superconducting cables. Table I summarizes the cable configurations for these three architectures.

#### 2. Analysis of the generated heat

There are two fundamentally different sources of heat: passive load and active load. Passive loads correspond to heat generated by thermal fluctuations at room temperature conducted down through cables with nonzero thermal conductivity. Passive heat is produced along the cables, due to their thermal resistivity, and within attenuators. In this analysis, we use the calculated passive heat per cable shown in Table II. Data are based on Table 2 from Ref. [55], considering the author's note that measured values might overestimate the exact passive heat. Passive heat from drive and flux lines differ due to different attenuator configurations. Optimal superconducting wire passive heat is taken from twisted-pair data in Fig. 1 of Ref. [55].

The total expected passive heat per qubit in each configuration is shown in Table III. We use the cable configuration from Table I in addition to the expected multiplexing in readout lines and the passive heat from the attenuators. For the QB configurations, we keep one drive line per ten qubits to account for the lines required for charging the cavity. This additional line adds only passive heat as the battery is charged prior to the computation.

TABLE I. Cable configuration for the analyzed architectures. The cable types are stainless steel coaxial cables (SS), niobium-titanium cables (NbTi), optimal superconducting cables (opt), and w. denotes with.

| Architecture | Drive | Flux | Feed | Output | Pump |
|---|---|---|---|---|---|
| Standard | SS | SS | SS | NbTi | SS |
| QB | SS | SS | SS | NbTi | SS |
| Standard w. SC cables | SS | Opt | SS | NbTi | SS |
| QB w. SC cables | SS | Opt | SS | NbTi | SS |

TABLE II. Calculated passive heat per cable type at *CP* and MXC stages. We use values from Table 2 in Ref. [55], averaged between measured and estimated passive heat. Optimal wire parameters are from Fig. 1 in Ref. [55].

| Cable type | *CP* passive load | MXC passive load |
|---|---|---|
| UT-085-SS-SS (Drive) | 365 nW | 8.5 nW |
| UT-085-SS-SS (Flux) | 270 nW | 17 nW |
| UT-085-NbTi (NbTi) | 240 nW | 11 nW |
| Optimal SC wire (opt) | 1 nW | 0.01 nW |

TABLE III. Total passive heat per qubit for each option, given the cable configuration above. The standard case is calculated through 1.25 SS drive cables, 1 SS flux, and 0.125 NbTi cables per qubit. The QB case is calculated through 0.35 SS drive cables, 1 SS flux, and 0.125 NbTi. The standard with SC cables case is calculated through 1.25 SS drive cables, 1 opt cable, and 0.125 NbTi cables per qubit. The QB with SC cables case is calculated through 0.35 SS drive cables, 1 opt cable, and 0.125 NbTi cables per qubit.

| Architecture | *CP* passive | MXC passive |
|---|---|---|
| Standard | $P_{\mathrm{p,CP}} = 756$ nW | $P_{\mathrm{p,MXC}} = 29$ nW |
| QB | $P_{\mathrm{p,CP}} = 427$ nW | $P_{\mathrm{p,MXC}} = 21.35$ nW |
| Standard w. SC cables | $P_{\mathrm{p,CP}} = 487$ nW | $P_{\mathrm{p,MXC}} = 12$ nW |
| QB w. SC cables | $P_{\mathrm{p,CP}} = 158$ nW | $P_{\mathrm{p,MXC}} = 4.36$ nW |

Active loads arise from the external (electromagnetic) control pulses, which dissipate along their path inside the cryogenic fridge, either within the cables of within attenuators. Attenuation is mandatory to reduce the incoming noise radiation field. In other words, attenuators are needed to block the high occupation of room-temperature photons to reach the quantum computer and modify the quantum state. As a result, the input powers of the drive signals in room temperature are significantly larger than the ones used to interact with the qubits. The active heat for each channel is given below.

*Drive channels*—These channels include 60-dB total attenuation ($10^6$ photon suppression), with 20 dB each at the MXC and $CP$. Reference [55] estimates average power per 20 ns $\pi$ pulse as $P_{\mathrm{avg}}(\pi) \approx -71$ dBm ($8 \times 10^{-11}$ W) and per $\pi/2$ pulse as $P_{\mathrm{avg}}(\pi/2) \approx -77$ dBm. Accounting for an average duty cycle of $D = 0.2$ per qubit (surface-code QEC + dynamical decoupling), the average active power per qubit becomes $P_{\mathrm{avg}} = D(P_{\mathrm{avg}}(\pi) + P_{\mathrm{avg}}(\pi/2))$, giving

$$P^{D}_{\mathrm{a,MXC}} = 10^2 P_{\mathrm{avg}} = 2.5\ \mathrm{nW},$$
$$P^{D}_{\mathrm{a,CP}} = 10^4 P_{\mathrm{avg}} = 250\ \mathrm{nW}.$$

In the QB configurations, drive lines used for battery charging do not create active heat since charging is performed before computation.

*Flux channels*—Dissipation arises mainly from dc bias currents setting qubit frequencies. Reference [55] gives worst-case estimates of 0.050 μW (MXC) and 0.140 μW ($CP$) per channel. Including around 1/6 additional overhead for entangling gates but reducing by a factor of 3 (with some magnetic shielding) gives

$$P^{F}_{\mathrm{a,MXC}} = 20\ \mathrm{nW}, \qquad P^{F}_{\mathrm{a,CP}} = 54\ \mathrm{nW}.$$

In the optimal SC cable case, flux active loads are negligible.

*Readout resonator feed lines*—Signals are about 10 times smaller than qubit drive channels. Assuming the same duty cycle but with one channel per eight qubits gives per-qubit powers

$$P^{RD}_{\mathrm{a,MXC}} = 30\ \mathrm{pW}, \qquad P^{RD}_{\mathrm{a,CP}} = 3\mathrm{nW}.$$

*Readout output channels*—These channels do not contribute active heat.

*Readout amplifier pump channels*—Pumping an amplifier requires around −60 dBm input with 10% duty cycle, attenuated by 50 dB total (10 dB in $CP$, 20 dB in MXC). With one pump per eight qubits, we find

$$P^{P}_{\mathrm{a,MXC}} = 2.5\ \mathrm{nW}, \qquad P^{P}_{\mathrm{a,CP}} = 25\ \mathrm{nW}.$$

TABLE IV. Total heat per qubit per configuration, combining passive and active loads.

| Architecture | $CP$ total | MXC total |
|---|---|---|
| Standard | 1024 nW | 52 nW |
| QB | 497 nW | 42 nW |
| Standard w. SC cables | 701 nW | 15.2 nW |
| QB w. SC cables | 173 nW | 5.6 nW |

TABLE V. Estimated maximum number of qubits per configuration, derived from total heat per qubit and fridge cooling capacities.

| Architecture | $CP$ qubits | MXC qubits | Limit |
|---|---|---|---|
| Standard | 976 | 657 | 657 |
| QB | 2011 | 808 | 808 |
| Standard w. SC cables | 1426 | 2226 | 1426 |
| QB w. SC cables | 5755 | 6033 | 5755 |

Table IV summarizes the total of both active and passive heat per qubit per configuration for each cryogenic stage. The total number of qubits per fridge is derived by taking these heating powers and comparing them to the state-of-the-art cryogenic fridge (Bluefors XLD1000sl) cooling powers of $P_{\mathrm{cool,MXC}} = 34$ μW and $P_{\mathrm{cool,CP}} = 1000$ μW. When taking the minimum in the number of potential qubits in the $CP$ and the MXC, we find the overall qubit limit of the cryogenic fridge per configuration, shown in Table V. In Fig. 9, we show the contribution of the different heat sources in the overall cooling budget. We note that, as opposed to the results in Ref. [55], the limit comes from the MXC and not the $CP$ stage since the current state-of-the-art cryogenic fridge has improved the $CP$ cooling power by a factor of 5 while the MXC cooling power was improved by a factor of 1.7.

### 3. Computation energy analysis

Power calculations shown in Fig. 7(c) are based on active heat dissipation at the $CP$, scaled by total attenuation up to room temperature. For a cycle duration of 1 μs, each gate contributes power according to the duty cycles described earlier. The flux channels add 10 dB, pump channels add 20 dB, and drive and readout drive channels add 20 dB relative to $CP$ power levels. For QB implementations, we add an energy overhead equivalent to 100 rounds of average drive power for a single qubit.

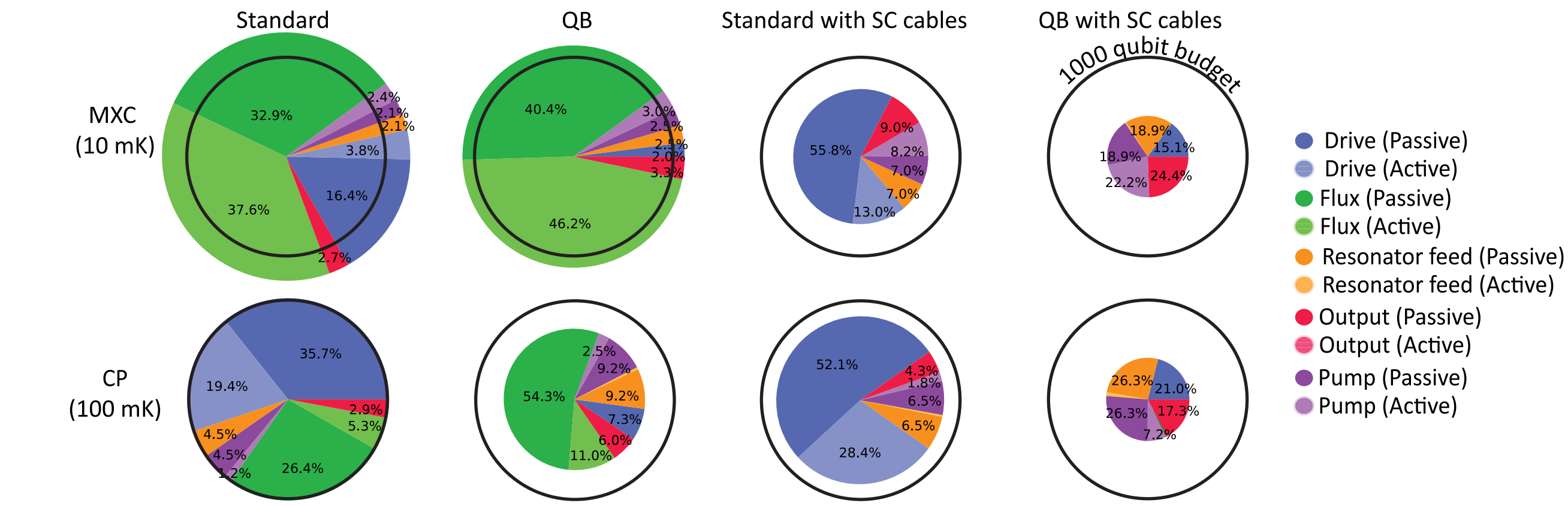

FIG. 9. Heat power source distribution for the examined computation architecture.

## APPENDIX C: CIRCUIT SIMULATION DETAILS

All quantum simulations were performed in Python using the QuTiP package. For the superextensive computation shown in Fig. 6 of the main text, we employed an extended Hilbert space that includes both cavity and qubit degrees of freedom. In contrast, all other simulations used a reduced Hilbert space of dimension $2^N$, corresponding to the cavity-qubit dressed states. In Fig. 6, all qubits were set to zero detuning, and the gate duration was chosen as the time maximizing fidelity with the target state $|1\rangle^{\otimes N}$. For Fig. 4, detuning values and durations were optimized for each gate in the circuit using SciPy's minimize function, with the cost function defined as the Hilbert-Schmidt distance between the implemented and target unitaries. In the simulations of Fig. 5, detuning values and gate durations were optimized to maximize the fidelity to the intermediate states in Fig. 5(c) and to the final state in Fig. 5(d). In these cases (and in Fig. 4), we allowed optimization across multiple gate steps. We found that providing initial guesses informed by analytical approximations significantly improved the convergence speed of the numerical optimization.